\documentclass[usenatbib,fleqn]{mn2e}
\pdfoutput=1 
\usepackage{amsmath, amssymb, wasysym, natbib, times, array}
\usepackage[pdftex]{graphicx}

\newcommand{\rcor}{\ensuremath{{r}_{\rm cor}}}

\newcommand{\rmag}{\ensuremath{{r}_{\rm mag}}}
\newcommand{\rhov}{\ensuremath{{\rho}{v}_{\rm p}}}

\newcommand{\flux}{\ensuremath{{\bf F}}}
\newcommand{\mdot}{\ensuremath{\dot M}}
\newcommand{\ldot}{\ensuremath{\dot L}}
\newcommand{\msun}{\ensuremath{M_\odot}}
\newcommand{\rsun}{\ensuremath{R_\odot}}
\newcommand{\comment}[1]{} 

\title[Propeller outflows from an MRI disc]{Propeller-driven outflows from an MRI disc}
\author[Lii et al.]{\parbox{\textwidth}{
Patrick S.~Lii$^{1}$\thanks{E-mail of corresponding author: \texttt{pslii@astro.cornell.edu}},
Marina M.~Romanova$^{1}$, 
Galina V.~Ustyugova$^{2}$,
Alexander V.~Koldoba$^{2}$,
Richard V.E.~Lovelace$^{1,3}$}\vspace{0.4cm}\\
\parbox{\textwidth}{ 
$^{1}$Department of Astronomy, Cornell University, Ithaca, NY 14853-6801\\
$^{2}$Keldysh Institute for Applied Mathematics, Moscow, Russia\\
$^{3}$Department of Applied and Engineering Physics, Cornell University, Ithaca, NY 14853-680}}

\begin{document}

\date{\today}
\pagerange{\pageref{firstpage}--\pageref{lastpage}} \pubyear{2012}

\maketitle

\label{firstpage}

\begin{abstract}
We present the results of axisymmetric simulations of MRI-driven accretion onto a rapidly rotating, magnetized star accreting in the propeller regime. The stellar magnetosphere corotates with the star, forming a centrifugal barrier at the disc-magnetosphere boundary which inhibits matter accretion onto the star. Instead, the disc matter accumulates at the disc-magnetosphere interface and slowly diffuses into the inner magnetosphere where it picks up angular momentum and is quickly ejected from the system as an outflow. Due to the interaction of the matter with the magnetosphere, this wind is discontinuous and is launched as discrete plasmoids. If the ejection rate is lower than the disc accretion rate, the matter accumulates at the disc-magnetosphere boundary faster than it can be ejected. In this case, accretion onto the star proceeds through the episodic accretion instability in which episodes of matter accumulation are followed by simultaneous accretion and ejection. During the accretion phase of this instability in which matter flows onto the star in funnel streams, we observe a corresponding rise in the outflow rate. Both the accretion and ejection processes observed in our simulations are highly non-stationary. The stars undergo strong spin-down due to the coupling of the stellar field with the disc and corona and we measure the spin-down timescales of around 1 Myr for a typical CTTS in the propeller regime.
\end{abstract}


\section{Introduction}
Young, newly formed protostars rotate rapidly and may be in the ``propeller'' regime of disc accretion in which the angular velocity of the stellar magnetosphere is larger than the angular velocity of the inner disc \citep{Illarionov1975}. Examples of other propelling stars include accreting fast-rotating neutron stars as well as white dwarfs in cataclysmic variables \citep{Stella1986, Cui1997, Alpar2001, Patruno2009}. In these stars, accretion in the inner disc is determined by the relative locations of the disc-magnetosphere boundary\footnote{Determined by the balance between the magnetic and matter pressures in the inner disc, so that the kinetic plasma parameter at the disc-magnetosphere boundary $\beta_1=(p+\rho v^2)/(B^2/{8\pi})\approx 1$.} \rmag\ and the corotation radius\footnote{The radius at which the angular velocity of the star $\Omega_*$ equals the local Keplerian angular velocity.} $\rcor=(GM_*/\Omega_*^2)^{1/3}$. If \rmag \textless \rcor, then the magnetosphere rotates slower than the inner disc and the disc is disrupted, causing the matter to accrete onto the star as a funnel flow \citep{Lamb1973, Elsner1977, Romanova2002}. In the opposite situation where \rmag \textgreater \rcor, the magnetosphere rotates more rapidly than the inner disc. In this case, the matter in the inner disc may acquire angular momentum from the magnetosphere, leading to super-Keplerian rotation of the matter which may in turn drive an outflow \citep{Lovelace1999}. This is known as the propeller regime of accretion.

The propeller regime has been studied analytically using magnetohydrodynamics \citep[MHD, e.g.][]{Davies1979, Li1997, Lovelace1999, Ikhsanov2002, Rappaport2004, Ekcsi2005} as well as numerically in MHD simulations \citep{Wang1985, Romanova2004, Romanova2005, Ustyugova2006, Romanova2009}. Numerical studies by \citet{Romanova2005} and \cite{Ustyugova2006} confirmed that propelling stars can launch strong outflows as predicted by \citet{Lovelace1999}. The outflows are found to have two main components: (1) a conical-shaped, centrifugally-driven magnetospheric wind into which much of the disc matter flows and (2) a high-velocity, magnetically driven and collimated axial jet into which a significant amount of energy and angular momentum flows. These previous simulations incorporated an $\alpha$-prescription \citep{Shakura1973} to model the viscosity and magnetic diffusivity of the disc plasma. Additionally, the simulations were performed on a spherical grid assuming top-bottom symmetry across the equatorial plane as well as axisymmetry about the rotation axis. More recently, one-sided outflows and opposite sided funnel flows were found in axisymmetric simulations of accretion of a viscous and diffusive $\alpha$-disc onto a rapidly rotating magnetized star for conditions where symmetry about the equatorial plane was not imposed \citep{Lovelace2010}.

Here, we study the problem of propeller outflows utilizing a new code which has a high spatial resolution in the region occupied by the disc. The higher resolution grid resolves the MHD turbulence generated by the magnetorotational instability \citep[MRI,][]{Balbus1991}. The MRI-driven turbulence causes accretion of the disc and for this reason we do not include either $\alpha$ viscosity or diffusivity coefficients, considering only an ideal MHD disc. Earlier, this code was used to study MRI-driven accretion onto slowly rotating stars \citep{Romanova2011}.

MRI-driven accretion has been studied in shearing boxes as well as in global simulations onto {\it non-magnetized} stars or black holes \citep[e.g.,][]{Hawley1995, Stone1996, Armitage1998, Gammie1998, Hawley2000, Hawley2001, Stone2001, Hawley2002, Beckwith2009}. Recently, our group was able to investigate MRI-driven accretion onto a {\it magnetized} star, first under axisymmetric conditions \citep{Romanova2011} and later in global 3D simulations \citep{Romanova2012}. It was shown that the disc matter flows onto the star through funnel streams \citep[as in][]{Romanova2002, Romanova2003, Romanova2004}.

In the present work, we model and study MRI-driven disc accretion onto a star in the propeller regime where $r_{\rm mag} > r_{\rm cor}$. In this regime we observe episodic funnel accretion which is accompanied by centrifugally-driven outflows launched from the disc-magnetosphere boundary. The aim of this work is to analyze the accretion and outflow processes in the propeller regime. In \S \ref{sec_model}, we discuss the numerical model and initial conditions of the simulations.  In \S \ref{sec_results}, we present an overview of the three models which we consider in this work; in \S \ref{subsec_outflows} we investigate the launching mechanisms; in \S \ref{subsec_spindown} we measure the spin-down rates for each of our models; and in \S \ref{sec_discussion}, we discuss the structure of the outflows and the application of our models to real systems.


\section{The numerical model} \label{sec_model}

\paragraph*{Code description:}
We use a Godunov-type numerical method with a five-wave Harten, Lax, and van Leer (HLL) Riemann solver similar to the HLLD solver developed by \citet{Miyoshi2005}. The MHD variables are calculated in four states bounded by five MHD discontinuities: the contact discontinuity, two Alfv\'en waves and two fast magnetosonic waves. Unlike \citet{Miyoshi2005}, our method solves an equation for the entropy instead of the full energy equation. This approximation is valid in cases such as ours where strong shocks are not present. In order to avoid non-physical solutions of the Riemann problem, we perform a correction procedure on the fast magnetosonic wave velocities in order to maintain the gap between these waves and the Alfv\'en waves which propagate behind the fast magnetosonic waves.  We ensure that the magnetic fields are divergence-free by introducing the $\phi$-component of the magnetic field potential which is calculated using the constrained transport scheme proposed by \citet{Gardiner2005}. To minimize the error in the Lorentz force, we split the magnetic field into the stellar dipole and calculated components, $B=B_{\rm dip} + B'$, omitting terms of the order $B_{\rm dip}^2$ which do not contribute to the Maxwellian stress tensor \citep{Tanaka1994}. No viscosity or diffusivity terms have been included in the MHD equations and hence we investigate only accretion driven by the resolved MRI-turbulence. Our code has been extensively tested and has been previously utilized to study MRI-turbulent accretion onto a magnetized star \citep[see tests and other details in][]{Romanova2011}.

\begin{figure}
\centering
\includegraphics[width=.45\textwidth]{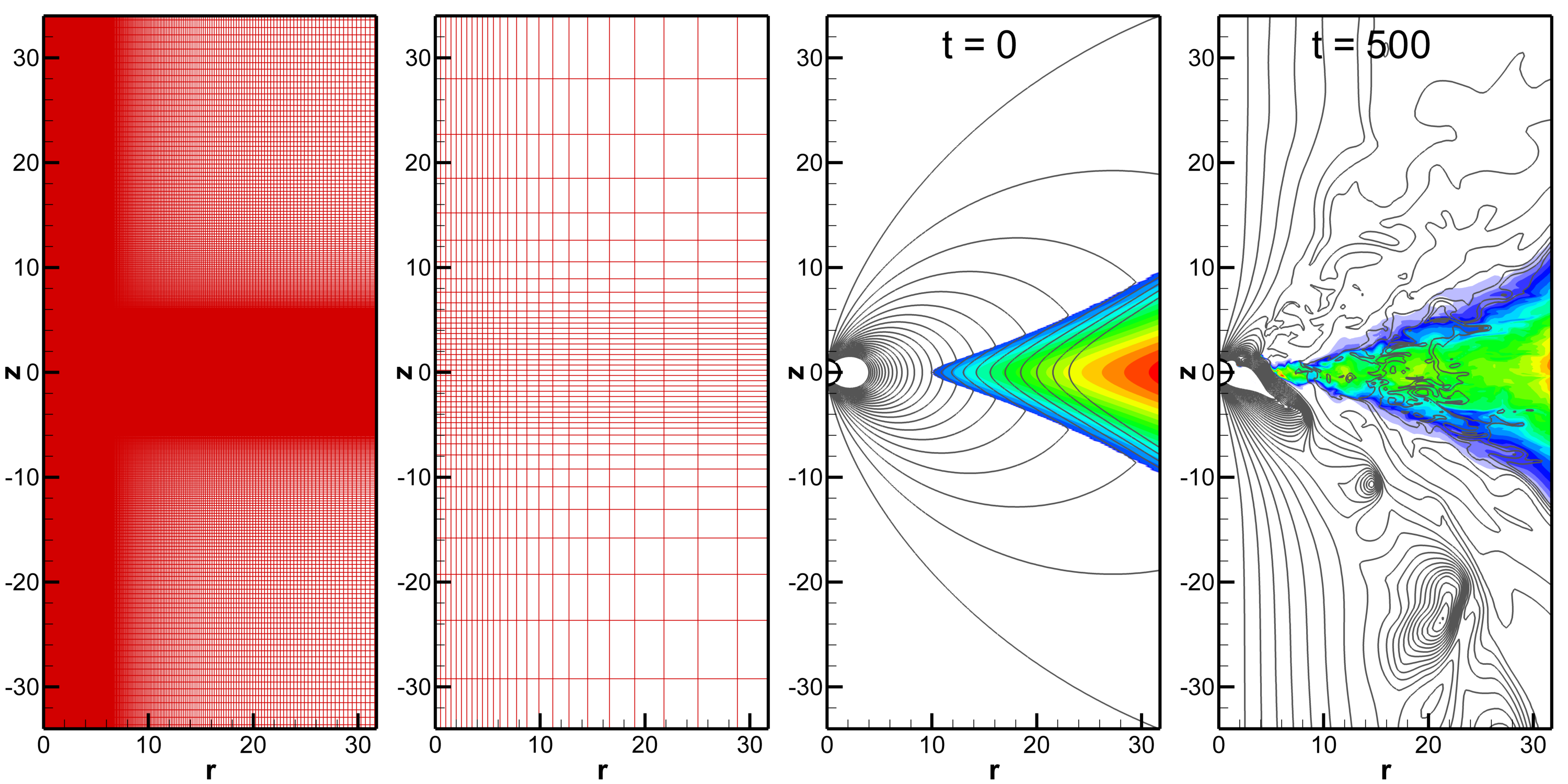}
\caption{{\bf The Grid.} From left to right: (1) grid used in simulations, (2) same grid but with 1/10th of the grid lines shown for clarity; (3) initial distribution of density (background) and magnetic flux (lines); (4) same but for time $t=500$.} \label{fig_grid}
\end{figure}

\paragraph*{Grid description:}
The axisymmetric grid uses cylindrical ($r$, $z$) coordinates with mesh compression towards the disc and towards the $z$-axis such that there are a larger number of cells in the disc plane and near the star (see Fig. \ref{fig_grid}, left panel). In the models presented here, we use a non-uniform grid with dimension $250 \times 432$ cells corresponding to a grid that is 33 by 70 stellar radii in size (0.31 AU by 0.65 AU if $R_* = 2 \rsun$). At $r=20$, the number of grid cells which cover the disc in the vertical direction is about 200.

\paragraph*{Reference units:} The simulations are performed in dimensionless units and are applicable to stars over a wide range of scales so long as the magnetospheric radius \rmag\ is not very large compared to the radius of the star $R_*$ (i.e., $\rmag = (4-6) R_*$). We choose the values of the stellar mass $M_*$, radius $R_*$, magnetic field $B_*$ and dimensionless magnetic moment $\tilde{\mu}$ = 10 and derive reference values using the initial values as a base. The derived reference units are as follows: length $R_0=R_*$, magnetic moment $\mu_0 = B_0 R_0^3$, magnetic field $B_0 = B_*/\tilde{\mu}$ with the dimensionless magnetic moment $\tilde{\mu}=10$ for our simulations, velocity $v_0=\sqrt{GM_*/R_0}$ (the Keplerian orbital velocity at the surface of the star), time period $P_0=2\pi R_0/v_0$ (the Keplerian orbital period at the surface of the star), angular velocity $\Omega_0 = v_0/R_0$, force per unit mass $f_0=v_0^2/R_0$, pressure $p_0 = B_0^2$, density $\rho_0=p_0/v_0^2$, temperature $T_0 = p_0/\rho_0$, accretion rate $\mdot_0=\rho_0 v_0 R_0^2$ and angular momentum flux $\ldot_0 = \mdot_0 v_0 R_0$. Table \ref{tab_ref} shows sample reference values for three different types of accreting stars: to apply the simulation results to a particular class of star, multiply the dimensionless value by the reference value.

For the remainder of this work, all values and variables are given in terms of the normalized units, except where they have explicitly assigned physical units. In all following plots, time and distance are always given in units of $P_0$ and $R_0$.

\begin{table}
\centering
\begin{tabular}{llll}
\hline \hline
          & CTTS    & White Dwarf        & Neutron Star    \\ \hline
{\bf initial} & & & \\ \hline
$M_*$ [\msun]  &      0.8 &        1 &      1.4 \\
$R_*$         &   2\rsun &  5000 km &    10 km \\
$B_*$ [G]   &      1000 &  $5 \times 10^5$ & $5 \times 10^8$ \\ \hline
{\bf derived} & & & \\ \hline
$R_0$ [cm]                  & $1.39\times 10^{11}$  & $5\times 10^{8}$    & $1\times 10^{6}$ \\
$B_0$ [G]                   & 100                 &  $5 \times 10^4$     &  $5 \times 10^7$ \\
$v_0$ [cm s$^{-1}$]          & $2.76 \times 10^{7}$ &  $5.15\times 10^{8}$  &  $1.36\times 10^{10}$ \\
$P_0$                       & 0.37 d              & 6.1 s               & 0.46 ms \\
$\Omega_0$ [$s^{-1}$]        & $1.99 \times 10^{-4}$   & 1.03                & $1.36 \times 10^4$ \\
$f_0$ [dy g$^{-1}$]      & $5.49 \times 10^{3}$  & $5.31 \times 10^{8}$ & $1.86 \times 10^{14}$ \\
$p_0$ [dy cm$^{-2}$]   & $1.0 \times 10^4$    & $2.5 \times 10^{9}$ & $2.5 \times 10^{15}$ \\
$\rho_0$ [g cm$^{-3}$]     & $1.31 \times 10^{-11}$ & $9.42 \times 10^{-9}$ & $1.35 \times 10^{-5}$ \\
$\mu_0$ [erg G$^{-1}$]	& $2.69 \times 10^{35}$ & $6.25 \times 10^{30}$ & $5 \times 10^{25}$ \\
$\mdot_0$ [\msun\ yr$^{-1}$] & $1.11 \times 10^{-7}$ & $1.92 \times 10^{-8}$ & $2.91 \times 10^{-9}$ \\
$\ldot_0$ [g cm$^2$ s$^{-2}$]            & $2.69 \times 10^{37}$ & $3.13 \times 10^{35}$ & $2.5 \times 10^{33}$ \\
\hline \hline
\end{tabular}
\caption{{\bf Reference units.} Calculated reference values for three different types of accreting stars. We choose the values of the stellar mass $M_*$, radius $R_*$, magnetic field $B_*$, and dimensionless magnetic moment $\tilde{\mu} = 10$ and derive the other reference values from these initial values.} \label{tab_ref}
\end{table}

\paragraph*{Initial conditions:} The initial conditions for the hydrodynamic variables are similar to those taken in our previous works \citep[e.g.][]{Romanova2002, Romanova2011} where the initial density and entropy distributions have been calculated from the force-balance of different forces which include gravity, centrifugal and pressure forces. The disc is dense and cold while the corona is about 1000 times hotter and of lower density.

\begin{table}
\centering
\begin{tabular}{llll}
\hline
\rcor & $\Omega_*$ & $P_*$ & $\tilde{\mu}$\\
\hline
1.3 & 0.67 & 9.31 & 10 \\
1.5 & 0.54 & 11.5 & 10 \\
2.0 & 0.35 & 17.8 & 10 \\
\hline
\end{tabular}
\caption{{\bf Models and model parameters.} In dimensionless units,  \rcor\ is the corotation radius, $\Omega_*$ is the angular velocity of the star, $P_*$ is the period of stellar rotation and $\tilde{\mu}$ is the dimensionless magnetic moment.} \label{tab_models}
\end{table}

We study three main cases of propelling stars in different rotational regimes with corotation radii \rcor\ =\ 1.3, 1.5, and 2 (see Table \ref{tab_models} for other parameters). At the start, the inner edge of the disc is placed at $r=10$ and the star rotates slowly with $\Omega_i$=0.032 (corresponding to \rcor = 10) such that the magnetosphere and inner disc are initially corotating with the same angular velocity. This condition helps to ensure that the magnetosphere and disc are initially in a near-equilibrium at the disc-magnetosphere boundary. From 0 to 100$P_0$, the star is gradually spun up from $\Omega_i$ to the final rapidly-rotating state with angular velocity $\Omega_*$ (given in Table \ref{tab_models}).

\paragraph*{Initial magnetic field configuration:} Initially, the disc is threaded with the dipole magnetic field of the star. We also add a small ``tapered'' poloidal field inside the disc (see third panel in Fig. \ref{fig_grid}) which is given by
$$
\Psi=\frac{B_0 r^2}{2}\cos\bigg(\pi\frac{z}{2h}\bigg), ~~
h=\sqrt{\bigg(\frac{GM_*}{\Phi_c(r)-E}\bigg)^2 - r^2},
$$
where $h$ is half-thickness of the disc, $\Phi_c(r)=k GM_*/r$ where $k$ is a Keplerian parameter\footnote{We take $k$ slightly greater than one to balance the disc pressure gradient (k=1+0.02).}, and $E$ is a constant of integration in the initial equilibrium equation \citep[see][]{Romanova2002, Romanova2011}. The magnetic field in the disc has the same polarity as the stellar field at the disc-magnetosphere boundary.

\paragraph*{Boundary Conditions:}
{\it Stellar surface:} all variables on the surface of the star have ``free'' boundary conditions such that $\partial (...)/\partial n=0$ along the entire surface. We prohibit the outflow of matter from the star (i.e. we do not allow for stellar winds) and adjust the matter velocity vectors to be parallel to the magnetic field vectors. This models the frozen-in condition on the star.

\noindent {\it Top and bottom boundaries:} all variables have free boundary conditions along the top and bottom boundaries. In addition, we implement outflow boundary conditions to prohibit matter from flowing back into the simulation region once it leaves.

\noindent {\it Outer side boundary:} the side boundary is divided into a ``disc region'' ($|z| < z_{\rm disc}$) and a ``coronal region" ($|z| > z_{\rm disc}$), with
$$
z_{\rm disc} = h(R_{\rm out}) = \sqrt{\left(\frac{GM_*}{\Phi_c(R_{\rm out})-E}\right)^2 - R_{\rm out}^2},
$$
where $R_{\rm out}$ is the external simulation radius. The matter along the disc boundary ($|z| < z_{\rm disc}$) is allowed to flow in with a small radial velocity
$$
v_r=-\delta \frac{3}{2}\frac{p}{\rho v_K(R_{\rm out})}, ~~\delta=0.02,
$$
and with a poloidal magnetic field corresponding to the calculated magnetic field at $r=R_{\rm out}$. The remainder of the variables are allowed to have free boundary conditions. The coronal boundary ($|z| > z_{\rm disc}$) has the same boundary conditions as the top and bottom boundaries.

\section{Results: outflows in the propeller regime} \label{sec_results}
We study three main cases of stars accreting in the propeller regime with different rotational periods corresponding to \rcor\ = 1.3, 1.5 and 2. For a representative T Tauri star ($M_*$ = 0.8 \msun, $R_*$ = 2 \rsun) these corotation radii imply  rotational periods of 3.4, 4.3, and 6.6 days, respectively. The dimensionless parameters applicable to other types of stars can be found in Table \ref{tab_models}. For stars accreting in the propeller regime, several processes are expected to occur: firstly, matter which diffuses into the rapidly rotating magnetosphere from the disc will be rapidly accelerated past the local Keplerian velocity and driven out as a propeller wind. In the regions above the star where the matter density is low, the magnetic field lines can easily inflate into the corona and the star may instead drive a magnetically dominated Poynting jet. Secondly, the rapidly rotating star should quickly spin down as angular momentum flows out from the star through the magnetic field lines connecting it to the corona and  disc. In this section, we study both of these effects: in \S \ref{subsec_outflows} we investigate the matter-dominated wind driven from the disc by the propelling star, in \S \ref{subsec_poynting} we discuss the magnetically-dominated Poynting component of the outflow, and in \S \ref{subsec_spindown} we measure the stellar spin down rate associated with the outflow in each of the models. 

\begin{figure*}
	\centering
		\includegraphics[width=\textwidth]{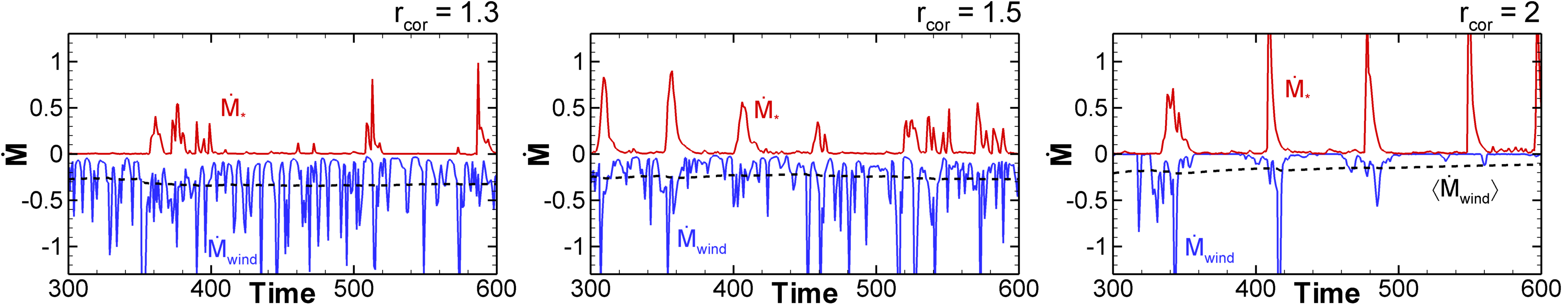}
		\caption{{\bf Mass fluxes in the three models.} The red lines show the accretion rate onto the star, $\mdot_*$, and the dark blue lines show the outflow rates $\mdot_{\rm wind}$. For reference, the time averaged outflow rates $\langle\mdot_{\rm wind}\rangle$ are also plotted as black dashes.}
	\label{fig_fluxes}
\end{figure*}

\subsection{Outflows in the propeller regime} \label{subsec_outflows}
The propeller outflow consists of two main components: a matter-dominated {\it propeller wind} launched from the disc-magnetosphere boundary and a magnetically-dominated {\it Poynting jet} which flows along the field lines connecting the star to the slower rotating corona. In this section, we discuss the matter dominated wind component. 

In order for the propelling star to eject matter from the disc, the magnetospheric velocity must exceed the Keplerian velocity at the disc-magnetosphere boundary, $v(r_{\rm mag}) > v_K(r_{\rm mag})$. If this condition is satisfied, the rapidly rotating magnetosphere acts as a centrifugal barrier and the matter in the disc is prevented from accreting onto the star. Instead, the matter accumulates at the disc-magnetosphere boundary and diffuses into the magnetosphere where it picks up angular momentum and is ejected as an outflow. The matter diffusion therefore determines the outflow rate and in Appendix \ref{appen_diff}, we further investigate the role of the diffusivity on the outflow rate in a special set of non-ideal MHD simulations. In the models, the outflow launched in this non-accreting state is non-stationary but proceeds continuously: when the matter is ejected, it inflates the field lines in the outer magnetosphere which expand and then quickly reconnect. This cyclic expansion-reconnection recurs on the dynamical timescale of the inner disc and the matter is ejected as discrete plasmoids similar to the clumpy winds observed in previous studies of propeller outflows \citep[e.g][]{Romanova2005, Ustyugova2006}. 

If the diffusion rate is low and the disc's accretion rate exceeds the outflow rate, the accreting matter is halted at the disc-magnetosphere boundary. Eventually, this accumulated matter accretes onto the propelling star through the episodic disc instability first described by \citet{Spruit1993} and studied further by \citet{DAngelo2010}. The basic mechanism behind the instability is as follows:
\begin{enumerate}
	\item The matter from the disc is blocked from accreting by the centrifugal barrier and accumulates at the disc-magnetosphere boundary, gradually compressing the stellar magnetosphere towards the star.
	\item The matter continues to compress the magnetosphere toward the star until the gravitational acceleration is larger than the centrifugal acceleration causing the matter to accrete onto the star as a funnel flow. 
	\item With the reservoir of matter depleted, the magnetosphere quickly re-expands out and shuts off the accretion, restarting the cycle anew.
\end{enumerate}
Episodic accretion results in a ``spiky'' accretion rate in which any accretion onto the star occurs in quasi-cyclic bursts. The accretion rate $\mdot_*$ in the top panels of Fig. \ref{fig_fluxes} show these characteristic bursts of matter in each of the three propeller models which we consider here. We measure the wind outflow rate, $\mdot_{\rm wind}$, by calculating the matter flux through a cylindrical surface centered on the star with radius $r=20$ and height $z=40$ and only measuring the fast component of the flow with $v_p \geq 0.2$. Figure \ref{fig_fluxes} shows that the outflows ($\mdot_{\rm wind}$) are highly non-stationary and vary strongly from model to model. In the simulations, these bursts of accretion tend to be accompanied by a burst of outflowing matter as well, as shown by $\mdot_{\rm wind}$. 

To get a clearer picture of the accretion and outflow rates, we calculate accretion rate time averages
\begin{equation}
\langle\mdot(t)\rangle = \frac{\int_{t_0}^{t} \mdot(t) dt}{\int_{t_0}^{t} dt} ~.
\label{eqn_timeavg}
\end{equation}
The time averaged $\langle\mdot_{\rm wind}\rangle$ is shown as a dashed line in Figure \ref{fig_fluxes}: the fastest rotating star has the largest overall outflow rate while the slowest rotating star has the smallest. We calculate accretion to ejection ratios $\langle\mdot_*\rangle / \langle\mdot_{\rm wind}\rangle$ of 0.144, 0.345, and 0.964 for the \rcor\ = 1.3, 1.5, 2 cases respectively. These values indicate that the faster rotating stars in the \rcor\ = 1.3 and 1.5 cases eject most of the accreting matter into an outflow before it ever reaches the surface of the star. In contrast, the weaker \rcor\ = 2 propeller does not launch the matter as efficiently and $\langle\mdot_*\rangle \approx \langle\mdot_{\rm wind}\rangle$ for this case.

\subsubsection*{Quasi-steady outflows in the non-accreting phase}
\begin{figure*}
	\centering
		\includegraphics[width=\textwidth]{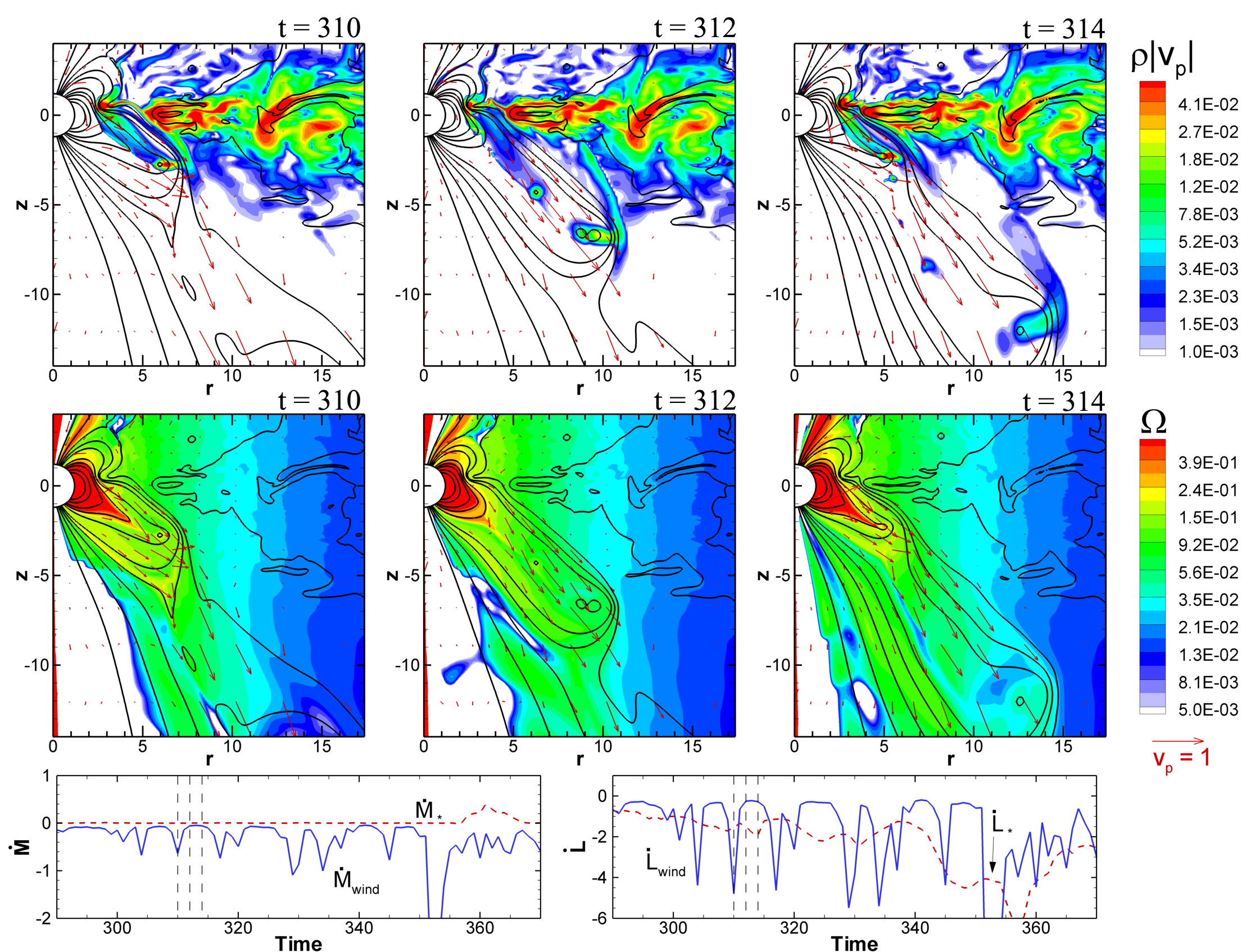}
		\caption{{\bf The non-accreting outflow mode.} {\it Top panels:} three snapshots of the \rcor=1.3 simulation region showing the plasmoids launched during the non-accreting phase of the episodic accretion cycle. Logarithmic maps of \rhov are plotted with field lines and velocity vectors superimposed. {\it Middle panels:} the angular frequency $\Omega$ of the matter at those same three moments in time. {\it Bottom panels:} matter and angular momentum fluxes onto the star and into the wind. The vertical dotted lines denote times corresponding to the three snapshots shown.}
	\label{fig_outflow}
\end{figure*}

During the accumulation phase of the accretion-ejection cycle, some of the matter at the disk-magnetosphere boundary diffuses through the disc-magnetosphere boundary into the inner magnetosphere where it picks up angular momentum and is ejected as a centrifugally-driven magnetospheric outflow. Figure \ref{fig_outflow} shows three time snapshots of these centrifugally launched outflows in the \rcor = 1.3 model. The top panels show the logarithmic poloidal matter flux $\rho |\mathbf{v}_p|$, poloidal velocity vectors $\mathbf{v}_p$, and the magnetic field lines near the star. In order for the matter to be driven into an outflow, it must overcome the local gravitational force as well as the magnetic tension in the magnetospheric field lines. The matter which flows out from the magnetosphere opens up the field lines, inflating the magnetosphere in the direction of the outflow (t = 310). As the accumulated matter flows away, the magnetic tension causes the field lines to reconnect (t = 312) and truncate the outflow into discrete lumps of matter which flow away at roughly 1.5 times the local escape velocity. As in previous studies of propeller outflows \citep[][]{Romanova2005}, we observe a half-opening angle of roughly 45$^\circ$ at the base of the propeller wind with evidence of gradual collimation at larger distances. The bottom panels of Figure \ref{fig_outflow} show snapshots of the angular velocity $\Omega$ alongside the poloidal velocity vectors $\mathbf{v}_p$ and stellar magnetic field lines. The large contrast in the angular velocities of the magnetosphere and inner disc is evident in the figure. Additionally, the reservoir of matter which builds up at the disc-magnetosphere boundary can be seen in the geometry of the magnetic field lines. 

An interesting feature of these propeller-driven outflows is that they are asymmetric. Previous studies of outflows from propelling stars have always assumed equatorial symmetry. Here, we see that one-sided centrifugally-driven outflows are launched even if the magnetosphere is initially top-bottom symmetric, similar to the one-sided magnetic outflows observed in simulations by \citet{Lovelace2010}.

\subsubsection{Outflows driven in the accreting phase} \label{subsubsec_funneloutflow}
\begin{figure*}
	\centering
		\includegraphics[width=\textwidth]{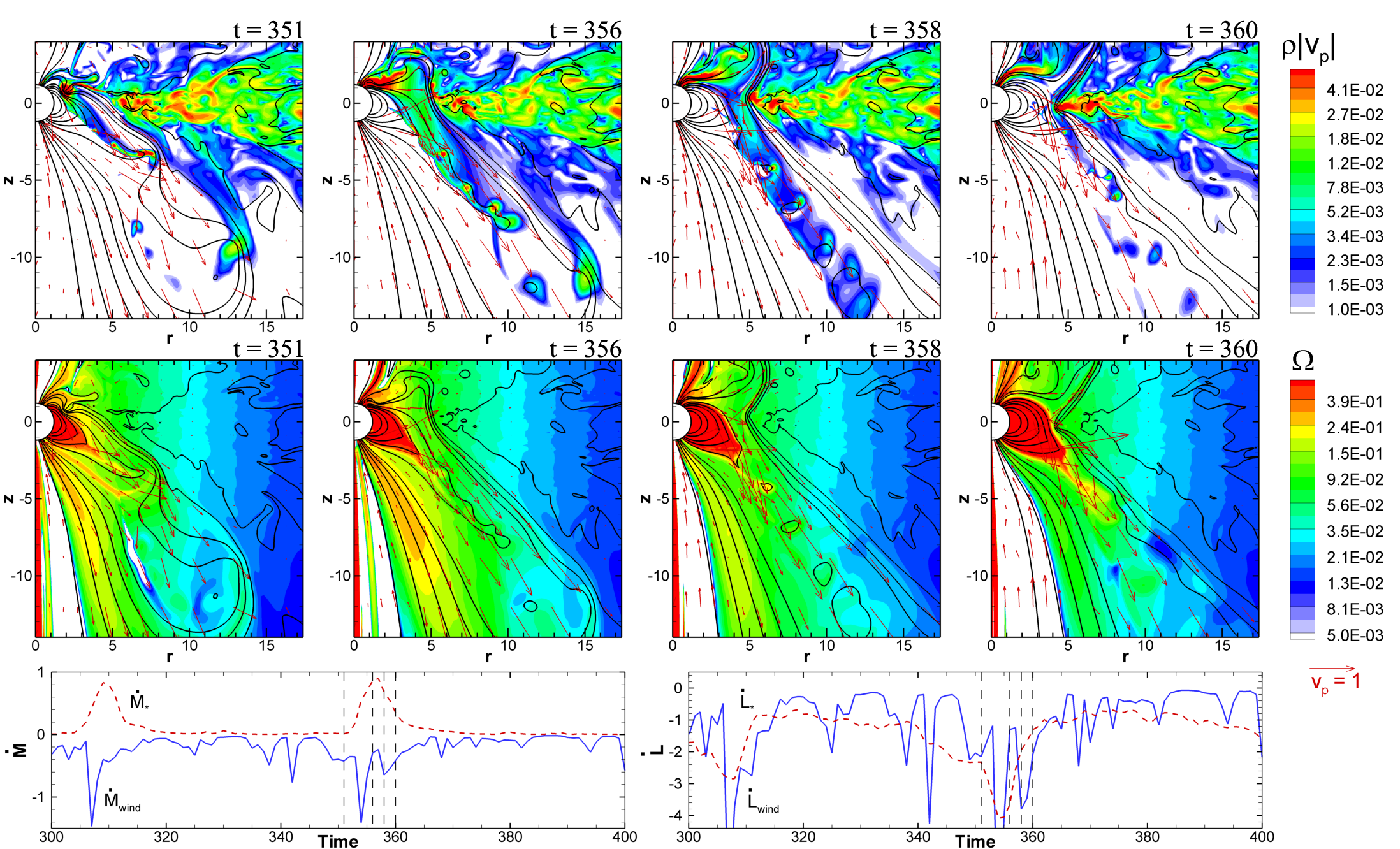}
		\caption{{\bf The accreting outflow mode.} {\it Top panels:} four time snapshots of the propeller outflow during an accretion event. A large amount of matter is ejected into the outflow when accumulated matter funnel flows around the magnetosphere from the disc onto the star. {\it Middle panels:} the angular frequency $\Omega$, velocity vectors and magnetic field lines at the same four moments in time. {\it Bottom panels:} the mass and angular momentum fluxes onto the star and into the wind. The vertical dashed lines correspond to the four time snapshots.}
	\label{fig_accoutflow}
\end{figure*}

In addition to the outflows in the non-accreting stage, the simulations show a stronger transient outflow associated with episodes of funnel flow accretion from the disc onto the star. Figure \ref{fig_accoutflow} shows four snapshots of one such accretion event in the \rcor = 1.3 model. At t = 351, the matter reservoir at the disc-magnetosphere interface has compressed the magnetosphere inward enough for accretion to proceed onto the star. By t = 356, a funnel flow has formed and the accretion and outflow rates have peaked: as the disc matter flows along the field lines toward the star, the contact area between the matter and magnetosphere is extended, permitting matter to diffuse through into the inner magnetosphere. There, the matter quickly picks up angular momentum from the field and is driven out into an outflow. Again, the diffusion of the matter across field lines plays a crucial role in loading the outflow (see Appendix \ref{appen_diff} for further discussion). Once the reservoir of matter compressing the magnetosphere is depleted by accretion onto the star, the magnetosphere  re-expands, halting the funnel flow of matter and reverting the outflow back to the initial weaker state. This episodic accretion event creates a short burst in the outflow rate occurring on a few dynamical timescales of the very inner disc. The stronger outflows in this phase tend to be more collimated than in the weaker, non-accreting phase with narrower half-opening angles ranging from 20--40$^{\circ}$ at the simulation boundary, depending on the amount of matter accreted onto the star.

In the simulations presented here, the disc accretion rate is high enough to compress the magnetosphere and drive the episodic accretion instability which leads to the spikes in the outflow rate. In stars with lower accretion rates or stronger fields, the magnetosphere may be very large compared to the disc thickness and accretion onto the star may be completely blocked. In these cases, the stronger outflows associated with the accretion may not be observed at all as funnel flows onto the star are prohibited.

\subsection{Poynting flux outflow} \label{subsec_poynting}
\begin{figure}
  \centering
  \includegraphics[width=.5\textwidth]{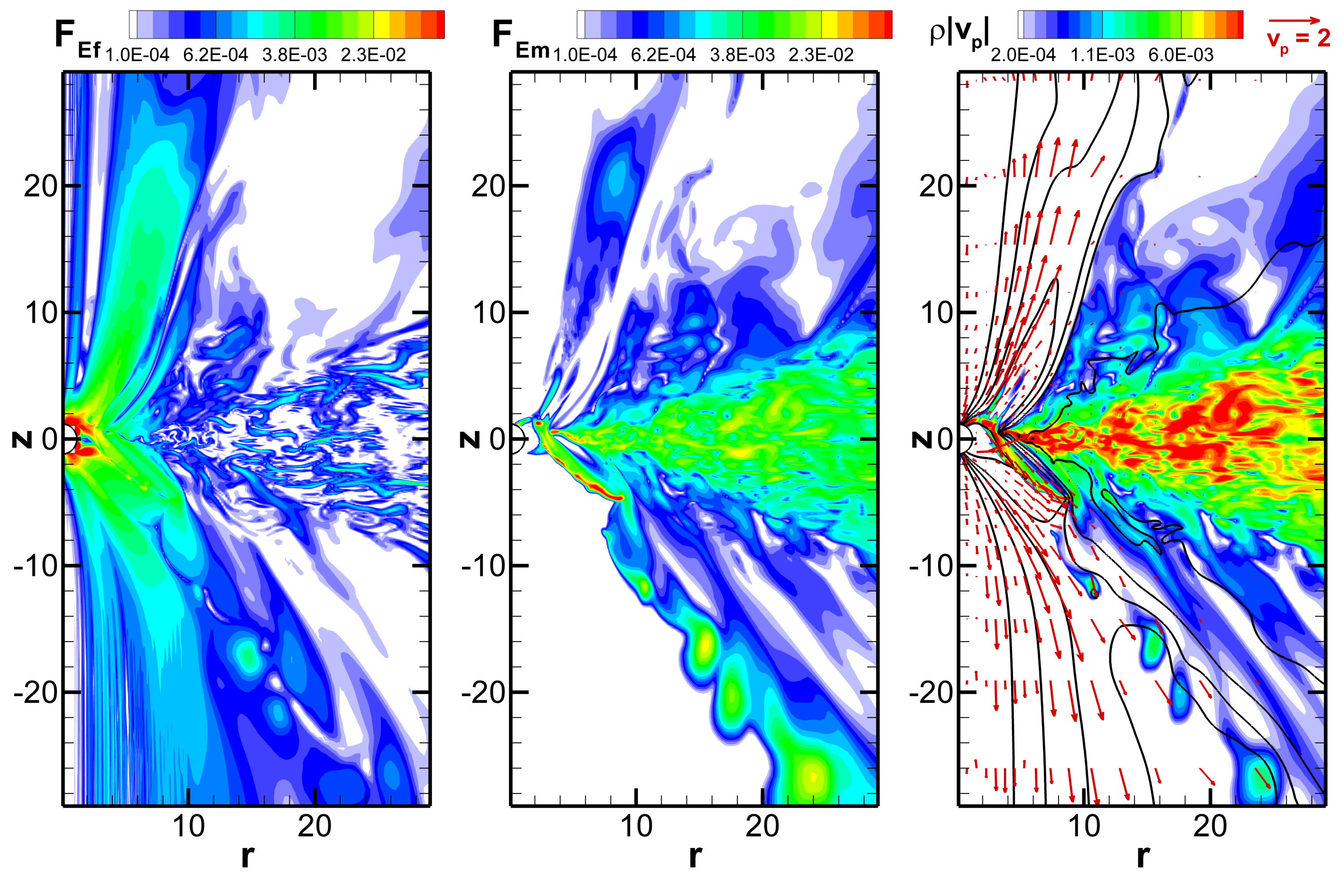}
  \caption{{\bf Energy flux densities.} {\it Left panel:} energy flux density carried by the magnetic fields. {\it Middle panel:} energy flux density carried by the matter. {\it Right panel:} logarithmic matter fluxes, poloidal field lines, and velocity vectors. All plots show the \rcor=1.3 model at t=400 during the accreting phase of the cycle.}
  \label{fig_energy}
\end{figure}

In addition to the matter-dominated propeller wind from the disc-magnetosphere boundary, we observe a magnetically dominated Poynting jet which flows along the helically wound field lines connecting the star with the slower rotating corona. In the jet, magnetic pressure accelerates the matter away from the star while the magnetic hoop stress simultaneously collimates it \citep{Ustyugova2000, Lovelace2002}. The jet is strongest during the ``accreting'' phase when the wind outflow is also the strongest because both phenomena are associated with the inflation and ejection of magnetic flux from the stellar magnetosphere. The magnetic Poynting outflow is well collimated with half-opening angles of 5-15$^\circ$ at the edge of the simulation region and the inner, strongly magnetically dominated regions of the jet exhibit more collimation than the outer regions of the outflow where there is more matter.

To study the energy outflow in the jet and wind, we measure the energy flux densities associated with the field
\begin{equation}
\mathbf{F}_{\rm Ef} = \frac{c}{4\pi} \mathbf{E}\times\mathbf{B} = 
\frac{|\mathbf{B}|^2 \mathbf{v_p} - (\mathbf{B}\cdot\mathbf{v}) \mathbf{B_p}}{4\pi}~,
\end{equation}
and the matter
\begin{equation}
\mathbf{F}_{\rm Em} = \left( \frac{\rho v^2}{2} + \frac{\gamma P}{\gamma-1} \right) \mathbf{v_p} \label{eqn_enrgflx}~,
\end{equation}
where $\mathbf{E} = -\mathbf{v} \times \mathbf{B}/c$ and $\gamma = 5/3$ for the ideal plasma in our models. Figure \ref{fig_energy} shows the energy fluxes at t = 400 during an episode of enhanced outflow into the wind: the leftmost panel shows the energy flux in the magnetic fields, highlighting the Poynting outflow from the star. As matter accretes, the enhanced magnetic outflow accelerates some of the accreting matter into a high velocity ($|{\mathbf v_p}| \approx$ 2), well-collimated jet. The middle panel shows the matter dominated wind also carries away substantial energy from the inner disc. The velocity vectors in the rightmost panel show that the accreting matter is accelerated to high velocities in both the magnetic and matter dominated regions of the flow. Even in the matter dominated components, there is a substantial amount of magnetic energy flux which helps to partially collimate the wind up to 20--40$^\circ$ at the edge of the simulation region. This magnetic flux will continue to collimate the flow at even larger distances.


\subsection{Propeller spin down} \label{subsec_spindown}

\begin{figure}
  \centering
  \includegraphics[width=.5\textwidth]{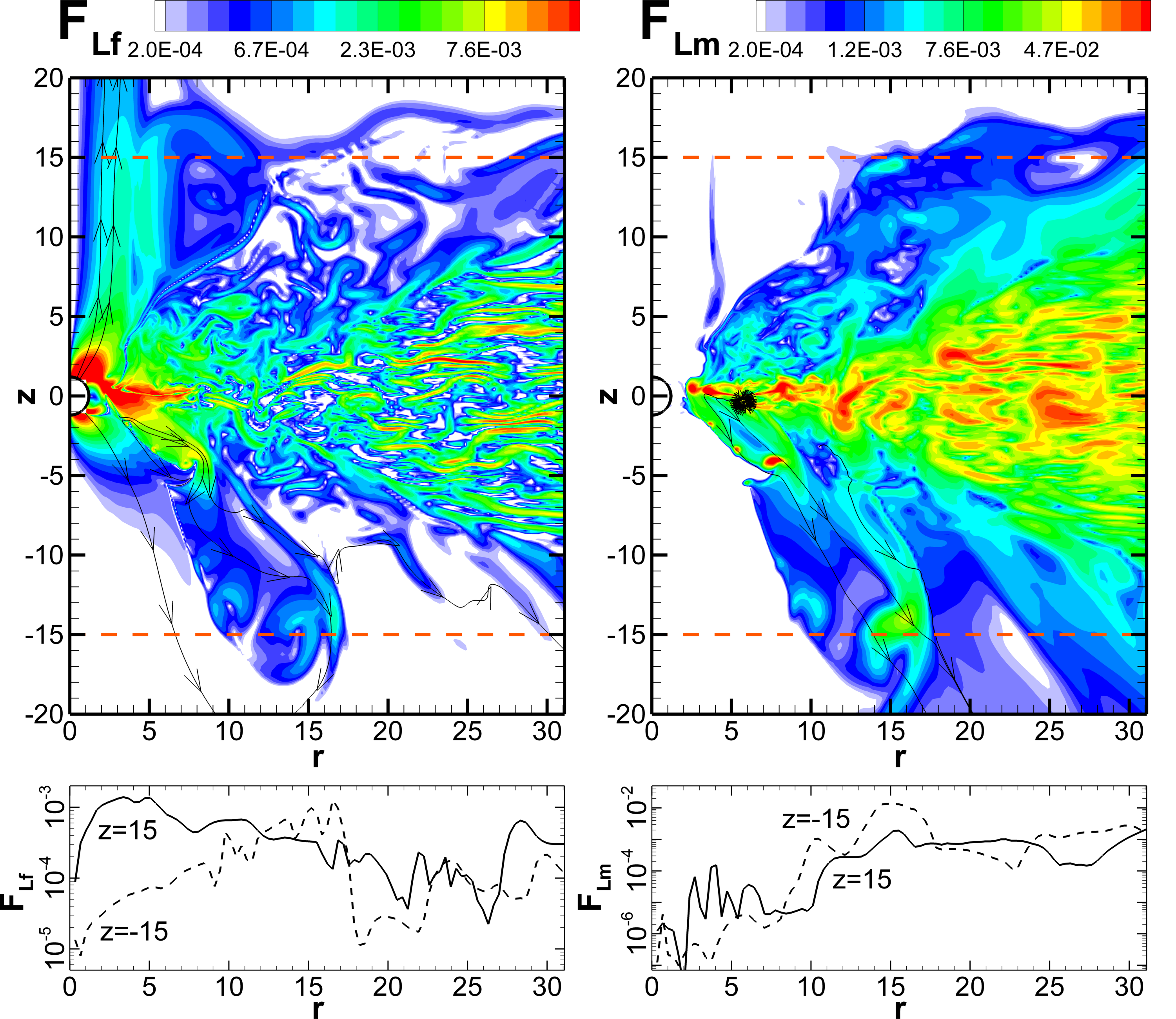}
  \caption{{\bf Angular momentum flux densities.} {\it Top left panel:} Angular momentum flux density carried by the magnetic fields. {\it Top right panel:} Angular momentum flux density carried by matter. All plots show the \rcor=1.3 model at t=315.} 
  \label{fig_2dangmom}
\end{figure}

The propelling star loses angular momentum through the stellar magnetic fields connecting the star to the slower rotating matter in the corona and disc. We are able to measure the angular momentum lost by the star by integrating the angular momentum flux densities over the stellar surface
\begin{equation}
\ldot = \ldot_m + \ldot_f = \int d{\mathbf S} \cdot (\flux_{\rm Lm}+\flux_{\rm Lf}) ~.
\end{equation}
Here the normal vector for the surface $d{\mathbf S}$ points inward towards the star and $\flux_{\rm Lm}$ and $\flux_{\rm Lf}$ are the angular momentum flux densities carried by the matter and magnetic field, given by
\begin{eqnarray} \label{eqn_angmom}
\flux_{\rm Lm} &=&  r \rho v_\phi  {\bf v}_p~,
\quad\quad \flux_{\rm Lf}= - r \frac{B_\phi {\mathbf B}_p}{4 \pi}~.
\end{eqnarray}
Figure \ref{fig_2dangmom} shows the angular momentum flux densities in the \rcor\ = 1.3 model at t = 315. 
The left panel shows the angular momentum being carried by the magnetic fields: a significant part of the angular momentum flux lies in the Poynting jet indicating that both the Poynting jet and propeller wind are important in spinning down the star. This spin-down torque is largely mediated by the magnetic field lines connecting it to the slower rotating corona and disc. The right panel, which shows the angular momentum transport through matter, highlights the matter stress induced by MRI in the disc as well as the substantial amount of angular momentum carried by the matter itself in the propeller-driven wind.

\begin{figure*}
  \centering
  \includegraphics[width=\textwidth]{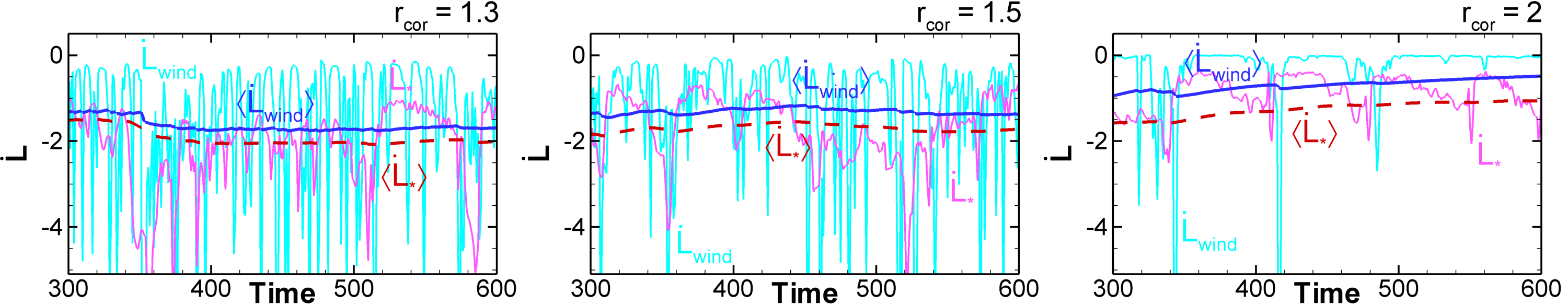}
  \caption{{\bf Angular momentum fluxes in the three models.} The dark red and blue lines show the time averaged angular  momentum flux onto the star and into the wind, respectively. The pink and light blue lines show the corresponding unaveraged fluxes onto the star and into the wind.}
  \label{fig_angmom}
\end{figure*}

To measure the long-term spin-down rate of the star, we time average the angular momentum fluxes 
$$\langle\ldot(t)\rangle = \frac{\int_{t_0}^{t} \ldot(t) dt}{\int_{t_0}^{t} dt},$$
and overplot the averages over the actual fluxes in Figure \ref{fig_angmom}. In agreement with the above analysis, these time averages indicate that the majority of the stellar angular momentum leaves in the outflow or as Poynting flux while the small fraction remaining is transported back outward through the disc by MRI. As shown by the unaveraged $\ldot_*$, the episodic nature of the accretion results in a torque which is episodic in nature as well. 

Using the time averaged angular momentum fluxes, we estimate the spin-down timescale for the stars in each of the three propelling stars. In dimensional units, the characteristic spin-down time is
\begin{equation}
t_{\rm sp} = \frac{L_*}{\ldot} = \frac{I \Omega_*}{\ldot} ~,
\end{equation}
where $I$ is the star's moment of inertia. Substituting in $\ldot = \tilde{\ldot} \ldot_0$ and $\Omega_* = \tilde{\Omega}_* \Omega_0$ (with tildes denoting the dimensionless values) and expanding to reference units, we have
\begin{eqnarray} \label{eqn_spindown}
t_{\rm sp} & = & \frac{I \tilde{\Omega}_* \Omega_0}{\tilde{\ldot} \ldot_0} \\
 & = & 9.14 \times 10^5 {\rm yr} 
\left(\frac{I}{1.2\times10^{55} {\rm g\ cm^2}}\right)
\left(\frac{M_*}{0.8 \msun}\right)^{\frac{1}{2}} \nonumber \\
 & & \left(\frac{R_*}{2\rsun}\right)^{-\frac{9}{2}}
\left(\frac{B_*}{1 {\rm kG}}\right)^{-2} 
\left(\frac{\tilde\mu}{10}\right)^2
\left(\frac{\tilde{\Omega}}{0.67}\right)
\left(\frac{\tilde{\ldot}}{2.0}\right)^{-1} ~, \nonumber
\end{eqnarray}
where we have plugged in values for a 0.8 \msun, 2 \rsun\ CTTS with $I \approx 1.23 \times 10^{55}$ g cm$^2$ and a 1 kG surface field (see Table \ref{tab_ref} for the other reference units). Using this, we can estimate the spin-down timescale for each of our cases:
\begin{enumerate}
\item \rcor\ = 1.3 
($\langle\tilde{\ldot}_{fs}\rangle \approx 2.0$), $t_{\rm sp, 1.3} \approx 9.65 \times 10^5$ yr.
\item \rcor\ = 1.5 
($\langle\tilde{\ldot}_{fs}\rangle \approx 1.7$), $t_{\rm sp, 1.5} \approx 9.10 \times 10^5$ yr.
\item \rcor = 2.0
($\langle\tilde{\ldot}_{fs}\rangle \approx 1.1$), $t_{\rm sp, 2} \approx 9.50 \times 10^5$ yr.
\end{enumerate}
The characteristic spin-down timescales are roughly the same for all three stars, hovering around one megayear. These characteristic timescales are roughly constant because the faster rotators have larger spin-down rates but also a larger amount of angular momentum. Table \ref{tab_summary} summarizes some of the dynamical properties of each of the three models. The derived spin-down timescales are consistent with measurements of CTTSs rotation periods which show that CTTSs older than 1 Myr are already slow rotators. As shown in Equation \ref{eqn_spindown}, the spin-down times depend on a large number of parameters---in particular, the stellar radius $R_*$, stellar magnetic field $B_*$, and the dimensionless magnetic moment $\tilde\mu$ which determines the size of the magnetosphere relative to the star. If the young star has a larger magnetosphere or radius than the given reference star, it will spin down even more rapidly that the 1 Myr timescale estimated here. 



\begin{table*}
\centering
\begin{tabular}{llllll}
\hline
\rcor [$R_*$] & $P_*$ [days] & $\langle\mdot_{\rm wind}\rangle / (\langle\mdot_{\rm wind}\rangle + \langle\mdot_*\rangle)$ & 
$\mdot_{\rm wind}$ [\msun\ yr$^{-1}$] & 
$t_{\rm sp}$ [Myr] & $\dot\Omega_*$ [s$^{-2}$] \\
\hline
1.3 & 3.4 & 0.87 & $3.61 \times 10^{-8}$ & 0.916 & 4.41 $\times 10^{-18}$  \\
1.5 & 4.3 & 0.74 & $1.03 \times 10^{-8}$ & 0.863 & 3.77 $\times 10^{-18}$ \\
2.0 & 6.6 & 0.51 & $1.21 \times 10^{-8}$ & 0.901 & 2.35 $\times 10^{-18}$ \\
\hline
\end{tabular}
\caption{{\bf Summary of values for CTTSs in the propeller regime.} These values are for a 0.8 \msun, 2\rsun CTTS with a 1 kG surface dipole field (see Table \ref{tab_ref} for other parameters). \rcor is the corotation radius, $P_*$ is the rotation period of the star, $\langle\mdot_{\rm wind}\rangle / (\langle\mdot_{\rm wind}\rangle + \langle\mdot_*\rangle)$ is the outflow ejection efficiency, $t_{sp}$ is the spin-down timescale of the star and $\dot\Omega_*$ is the stellar spin-down rate.}
\label{tab_summary}
\end{table*}

\section{Discussion} \label{sec_discussion}

\subsection{Propelling CTTSs}
Newly formed Class 0 T Tauri stars are expected to be in a state of rapid rotation and may have strong magnetic fields which  interact with their accretion discs. For this reason, propeller-driven outflows are expected in this  early stage of stellar evolution. However, in this phase, the nascent stars are enshrouded by an optically-thick layer of gas and dust, preventing direct observations of the inner launching region. Instead, observations of these young stars show Herbig-Haro jets emerging from a dense optically-thick proplyd which hides the star. One of the possible origins of these observed outflows is the propeller mechanism discussed in this work which channels substantial amounts of matter and energy into a well-collimated  outflow. 

In later stages of protostellar evolution (after $\sim 10^6$ yrs), the thick cloud of gas and dust enshrouding the star clears and the young star becomes accessible to direct observation. At this stage, measurements show the stars are relatively slowly rotating, with periods $\sim$2-10 days despite the fact that the stars continue to gravitationally contract. This is the Classical T Tauri star stage of the evolution (CTTSs, or Class II stars). The CTTSs show outflows which are much weaker than at the earlier stages of evolution. Propeller outflows will be present in this stage during episodes of low accretion when the stellar magnetospheric radius exceeds the corotation radius of the star. 

One such example of a propelling CTTS is the well studied Class II T Tauri star AA Tau, with inferred mass, radius and age of $0.7\msun$, $2\rsun$ and 1.5 Myr. Spectropolarimetric measurements by \citet{Donati2010} indicate that its magnetic field is dominated by a 2-3 kG dipole component inclined at $20^\circ$ to the rotation axis of the star. This inclined dipole creates a warp in the inner disc which corotates with the star, periodically eclipsing it every 8.22 days, indicating that the star has a corotation radius $\rcor = 7.6 R_*$ \citep{Bouvier2007}. Based on these parameters and measurements of the accretion rate ($10^{-9.2} \msun$/yr on average over their observation epochs), \citet{Donati2010} showed that AA Tau is well within the propeller regime of accretion with $\rmag \approx 2\ \rcor$. As the magnetospheric radius lies well beyond corotation, AA Tau is a prime candidate for launching propeller-driven outflows. Imaging studies of the AA Tau jet by \citet{Cox2013} reveal a clumpy high-velocity outflow with an unusually wide half-opening angle of $17^\circ$ within 2'' of the star. In the same study, a possible counterjet was suggested, but not confirmed. 

The microjets observed in young T Tauri stars are known to be well collimated at distances greater than 10 AU from the star. Observations suggest that they have a layered ``onion-skin'' structure, with a well-collimated, high-velocity central component surrounded by several less collimated, lower velocity layers. The well-collimated central components typically have half-opening angles of 2--12$^{\circ}$---narrower than the AA Tau jet and narrower than the 20-40$^\circ$ wind that we observe in our models \citep[e.g.,][]{Dougados2000, Agra-Amboage2011}. In order for the propeller wind to explain the jet phenomenon in T Tauri stars, a larger-scale collimation mechanism must be present. The most likely scenario is that the propeller wind continues to be self-collimated beyond the edge of our simulation region. Despite being matter dominated, the propeller wind carries a substantial amount of magnetic energy which is responsible for the partial collimation observed in our models (see \S \ref{subsec_poynting}). This magnetic flux will continue to collimate the flow at larger distances, resulting in a  well-collimated propeller outflow far from the star. It may also be possible for the propeller wind to be confined by the ram pressure of a slower moving larger-scale disc-wind or by the magnetic pressure of large scale poloidal fields threading the accretion disc, but it is not clear if these mechanisms are important as the magnetic collimation is already sufficient \citep{Frank1996, Matt2003}. A collimated propeller outflow may be responsible for the central medium/high-velocity components observed in T Tauri jets while the lower velocity outer layers may be due to a extended magneto-centrifugal disc-wind launched at larger distances. Observationally, it is not at all clear if the microjet in AA Tau is in fact propeller-driven, but the outflow velocity and unusually wide opening angle may be explained by a propeller wind which has undergone weak collimation.

\subsection{The one-sided propeller wind}

The geometry of the stellar magnetosphere plays a key role in the launching of the propeller outflow. In our axisymmetric models, the aligned dipole is initially equatorially symmetric. However, as soon as the matter from the disc reaches the magnetosphere, the stellar dipole is deflected to one side and this initial top-bottom symmetry is broken. This asymmetry helps to drive the one-sided outflows observed in this work as it provides a preferential direction for the matter accretion and ejection. For our simple dipole magnetosphere, we observe occasional ``flip-flops'' where the direction of the outflow switches sides. \citet{Lovelace2010} studied one-sided outflows driven by stars with both dipole and quadrupole components finding that an aligned, purely quadrupolar field could drive top-bottom symmetric outflows. However, the addition of even a small dipole component to the field caused one-sided outflows to be driven from the disc-magnetosphere boundary. As in the simulations presented here, \citet{Lovelace2010} also found that a pure dipolar field drives one-sided outflows which can occasionally flip sides.


The presence of a complex magnetic topology on the star may also help to drive one-sided outflows. Most CTTSs have fields which are inclined relative to the rotation axis---a configuration which is intrinsically non-axisymmetric and cannot be captured by our code. Additionally, several CTTSs are known where the higher order components are comparable to or stronger than the dipole components \citep[e.g. BP Tauri, V2129 Oph,][]{Donati2008,Donati2011a}. While the dipole may dominate the dynamics at larger distances in the disc, these higher order components play a major role in the dynamics near the star. Additionally, axisymmetric simulations do not capture cases such as the one where the dipole field on the western hemisphere of the star is deflected in the opposite direction than the field on the eastern hemisphere. However, all of these effects would likely enhance the one-sided nature of the outflow as they contribute to the asymmetry of the magnetosphere. Further MHD simulations in full 3D are necessary to study these magnetospheric dynamics.

\paragraph*{Weak propellers:}
It is possible for an accreting star to be in the weak propeller regime of accretion where the star rotates faster than the inner disc, but not fast enough to drive an outflow. This occurs if the centrifugal force at the disc-magnetosphere boundary is insufficient to overcome both the star's gravity and the magnetic tension opposing field line inflation. In this case some matter may still be ejected, but then it falls back onto the disc at larger distances from the star instead of flowing out of the system. In this sense, a weakly propelling star may continuously recycle the inner disc, causing the disc to evolve into a trapped state in which the accretion onto the star occurs in rare episodic bursts \citep{DAngelo2012}. In the weak propeller regime, the disc may also form transient funnel flows and the outflow may be launched as magnetospheric ejections \citep{Zanni2013}. However, if the propeller effect is especially weak, no matter may be ejected at all \citep[see also][]{Romanova2004}.

\subsection{The effect of the MRI}
In prior works \citep[e.g.][]{Romanova2005, Ustyugova2006, Romanova2009}, propeller-driven outflows were observed in an $\alpha$-disc where both the viscosity and diffusivity were modeled with numerical terms proportional to the \citet{Shakura1973} $\alpha$ parameter. In these prior simulations, relatively large viscosity coefficients of $\alpha_v$ = 0.1--0.3 were taken in the main models. Additionally, a large $\alpha$ diffusivity of $\alpha_d$ = 0.1 was taken in order to facilitate the penetration of the disc matter into the magnetosphere.

In the current work, we study the propeller outflow phenomenon in the case of a more physical MRI-driven disc which accretes due to the MRI-driven angular momentum transport. In contrast to the previous studies, no viscosity or diffusivity terms were added to the MHD equations, resulting in a more realistic model of the disc accretion. However, the MRI does not provide a source of diffusivity near the magnetically-dominated magnetosphere and the inner disc matter only penetrates the magnetospheric field lines due to the very small numerical diffusivity in our MHD code. Nonetheless, our simulations show that the propeller-driven outflows persist even with the small effective diffusivity. In actuality, the diffusivity may be much larger in nature due to 3D instabilities such as the Rayleigh-Taylor and Kelvin-Helmholtz instability which act at the disc-magnetosphere boundaries \citep{Romanova2008, Kulkarni2008}. 

In Appendix \ref{appen_diff}, we study the effects of a larger diffusivity on the propeller outflow by applying an $\alpha$-diffusivity to the matter near the disc-magnetosphere boundary. The larger diffusivity generally increases the outflow rate, but only by a factor of two even in the most diffusive cases. The dynamics of the disc-magnetosphere boundary is poorly understood and hence, the effective diffusivity in the inner-disc is also poorly known; however, the simulations show that the diffusivity does not strongly affect the physics of the outflow.

\section{Conclusions}

In this work, we have used high-resolution simulations to study propeller-driven outflows from a realistic disc which accretes due to MRI driven turbulence. In the propeller regime of accretion, the star (and its magnetosphere) rotates faster than the Keplerian angular velocity of the inner accretion disc, resulting in a large centrifugal barrier at the disc-magnetosphere boundary. The main conclusions of this work are as follows:

\begin{enumerate}
\item Stars accreting in the propeller regime can launch centrifugally driven outflows. The rapidly-rotating magnetosphere creates a large centrifugal barrier which blocks matter from accreting onto the star. Instead, the accreting matter accumulates at the disc-magnetosphere boundary over time. Some of this matter diffuses through the outer  magnetosphere into the more rapidly rotating inner regions where it picks up angular momentum from the field lines. If the centrifugal force is sufficiently large, the matter becomes super-Keplerian and inflates the stellar field, creating a path for the matter to flow out of the system. As the matter flows away some of the field lines reconnect, releasing magnetic flux and matter from the magnetosphere and truncating the outflow into discrete plasmoids. This matter dominated propeller wind is ejected with an half-opening angle of 45$^\circ$ and gradually collimated at larger distances.

The remainder of the matter which does not diffuse into the magnetosphere instead undergoes an episodic accretion-ejection cycle. As the matter accumulates at the disc-magnetosphere boundary, the magnetosphere is slowly compressed towards the star. Eventually, the gravitational force becomes dominant and the matter deflects the magnetosphere to one side and accretes onto the star as a magnetospheric funnel flow. A significant fraction of this accreting matter is driven into an outflow, resulting in an enhanced outflow rate. Once the accumulated matter is depleted, the stellar magnetosphere quickly re-expands to halt the accretion onto the star. The enhanced outflow results in the ejection of a larger amount of magnetic flux which helps to collimate the matter, giving a narrower half-opening angle (compared to the non-accreting phase) of 20--40$^\circ$ at the edge of the simulation region. 

\item The propeller outflow has two main components: a high density, medium velocity, matter dominated wind and a low density, high velocity, magnetically dominated Poynting jet. The Poynting jet is magnetically accelerated and collimated by the helically wound magnetic field lines connecting the star to the corona. It carries substantial amounts of magnetic energy and angular momentum away from the star and has a half-opening angle of 5-15$^\circ$ at the edge of the simulation region. The slower moving, matter dominated propeller wind is launched from the disc-magnetosphere boundary by the centrifugal force and helps to remove matter (and angular momentum) from the disc. Despite being matter dominated, the propeller wind also carries a substantial amount of magnetic flux which acts to weakly collimate the outflow. We observe weak collimation in our simulation region, but the magnetic hoop-stress will continue to collimate the flow resulting in a well-collimated jet at large distances from the star.

\item Stars in the propeller regime experience a spin-down torque through the magnetic field lines connecting the star to the slower rotating disc and corona and lose energy through the Poynting jet and propeller wind. The more rapidly rotating stars have higher spin-down rates and in all three of our models, we measure characteristic spin-down timescales of around 1 Myr for a typical CTTS with a strong 1 kG surface field (other parameters can be found in Table \ref{tab_ref}).
\end{enumerate}

\section*{Acknowledgments}
Resources supporting this work were provided by the NASA High-End Computing (HEC) Program through the NASA Advanced Supercomputing (NAS) Division at Ames Research Center and the NASA Center for Computational Sciences (NCCS) at Goddard Space Flight Center. The research was supported by NASA grants NNX10AF63G, NNX11AF33G and NSF grants AST-1008636 and AST-1211318. G. Ustyugova and A. Koldoba were supported in part by FAP-14.B37.21.0915, SS-1434.2012.2, and RFBR 12-01-00606-a.


\bibliographystyle{mn2e}
\bibliography{mn-jour,propeller}

\appendix

\section{Diffusivity and outflows} \label{appen_diff}
The MRI turbulence provides a mechanism for angular momentum transport in the disc and permits inward accretion from larger distances. However, it does not act as a source of diffusivity at the disc-magnetosphere boundary \citep{Romanova2011}. In spite of the MRI-driven turbulence, the accretion disc is matter-dominated and its dynamics are similar to a non-magnetized disc. The inner disc rotates at the Keplerian velocity whereas the magnetosphere is super-Keplerian and they do not interact unless there is a mechanism for diffusivity which provides penetration of the disc matter through the magnetosphere.

\begin{figure*}
\centering
\includegraphics[width=.7\textwidth]{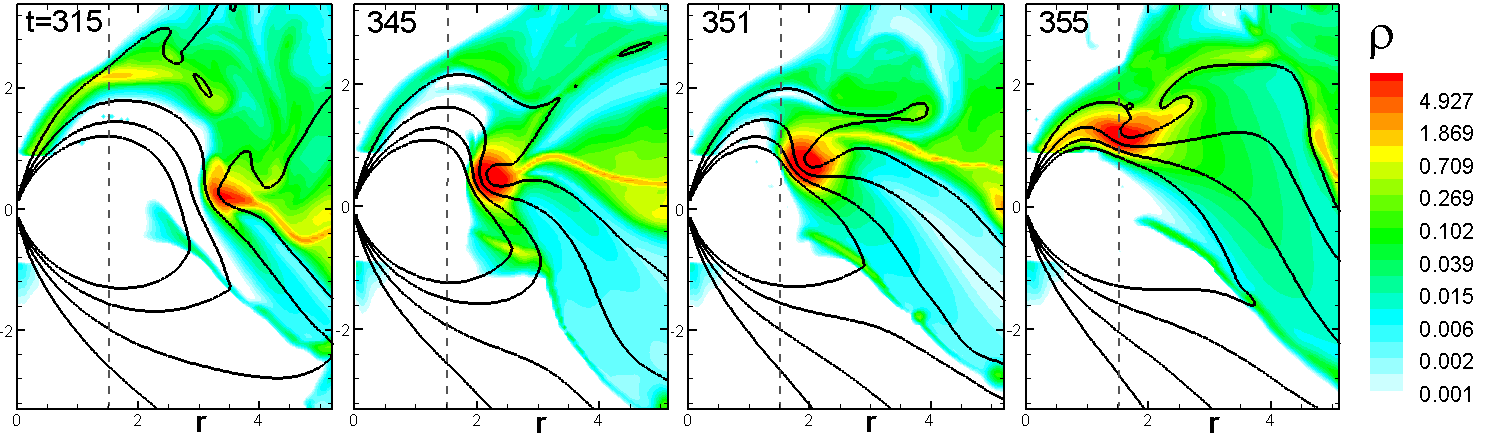}
\caption{{\bf Diffusive penetration through the magnetosphere.} The snapshots show the gradual diffusive penetration of the disc matter through several field lines of the outer magnetosphere. The color background shows the logarithmic density distribution with red indicating the maximum density and light-cyan indicating the smallest density in the disc. There is also an even lower density corona which is shown in white.} \label{fig_diff4}
\end{figure*}

\begin{table*}
\centering
\begin{tabular}{|c|l|lll|lll|} \hline
\multicolumn{1}{|c}{} & \multicolumn{1}{|c}{base case} & \multicolumn{3}{|c}{set 1: low $\rho_d$} & \multicolumn{3}{|c|}{set 2: high $\rho_d$} \\ \hline
$\alpha_d$ & 0.0     & 0.01       & 0.1 & 1.0   &  0.1   & 0.1 &1  \\[4pt]
$\rho_d$ & 0.0     & 0.01       & 0.01        & 0.01  &  1     & 5 & 5\\[4pt]
$\langle{\mdot}_{\rm wind}\rangle$ & 0.27    & 0.26       &0.4          & 0.34  &  0.34  & 0.16 & 0.24\\[4pt]
$\langle{\mdot}_*\rangle$ & 0.09    & 0.11       &0.08         & 0.05  &  0.12  & 0.08 & 0.07 \\[4pt]
$\frac{\langle{\mdot}_{\rm wind}\rangle}{\langle{\mdot}_{\rm wind}\rangle + \langle{\mdot}_*\rangle}$     & 0.74    & 0.70       &0.83         & 0.86  &   0.74 & 0.67   & 0.77\\[4pt]
$\langle{{\ldot}_*}\rangle$ & 1.7     & 2.0        & 2.7         & 2.25  &   2.0  & 1.70 & 1.75\\ \hline
\end{tabular}

\caption{{\bf Summary of results for the diffusive runs.} All runs were performed for \rcor = 1.5.
$\alpha_d$ is the diffusivity parameter, $\rho_d$ is the threshold density for diffusivity, $\langle{\dot M}_{\rm wind}\rangle$ and $\langle{\mdot}_*\rangle$ are time-averaged values of matter flux to the wind and to the star, respectively. 
The ratio $\langle{\mdot_{\rm wind}\rangle}/(\langle\mdot_{\rm wind}\rangle+\langle\mdot_*\rangle)$ is the outflow ejection efficiency of the propeller. 
$\langle\ldot_*\rangle$ is the  angular momentum flux from the surface of the star.}
\label{tab_diffusivity}
\end{table*}

In the ideal MHD simulations presented in the main text of this work, the matter in the inner disc gradually penetrates through the external magnetosphere due to numerical diffusivity.  Fig. \ref{fig_diff4} shows several snapshots of the accumulation of matter in the inner disc, the slow diffusive penetration through the external field lines of the magnetosphere, and the resulting outflow launched along the inflated stellar field lines. In our ideal simulations, the numerical diffusivity is low because the grid resolution is very high near the disk-magnetosphere boundary. In real systems, diffusive penetration is determined by the magnetic Reynolds number:
\begin{equation}
R_m=\frac{v {\Delta r}}{\eta_m} ~,
\end{equation}
where $v$ is the radial velocity of the inner disc matter, $\Delta r$ is the characteristic size of compressed layer of the magnetosphere through which the inner disc matter should diffuse, and $\eta_m$ is the coefficient of the magnetic diffusivity. At the disc-magnetosphere boundary of the propelling star, the inner disc is almost stopped by the magnetosphere and therefore the radial inward velocity $v$ is small. Additionally, the magnetosphere is compressed by the inner disc meaning that the characteristic scale $\Delta r$  is also small, in particular during episodes of matter accumulation. Therefore, the condition ${\rm Re_m}<1$, which is necessary for diffusive penetration, is satisfied even at small $\eta_m$ as we observe in our simulations.

The diffusivity of the stellar magnetosphere is one of the most poorly understood properties of young stars and may be larger than the numerical diffusivity present in the code. 
One of the main limitations of our axisymmetric simulations is that we cannot directly simulate mechanisms such as the Rayleigh-Taylor and Kelvin-Helmholtz instabilities which may be responsible for the mixing of the matter and field at the disc-magnetosphere boundary \citep{Arons1976}. In 3D, these mechanisms may serve as the primary source of diffusivity in the magnetosphere. The Kelvin-Helmholtz instability may be important because the inner disc matter rotates slower that the external magnetosphere. In the propeller regime, the Rayleigh-Taylor instability cannot act in its usual form because the presence of an inward pointing effective gravity is required \citep{Spruit1995, Kulkarni2008}. However, it may operate in the opposite direction, driven by the outward pointing centrifugal force. The enhancement of diffusivity can strongly influence the rate of accretion and outflows in the propeller regime.

To investigate the role of  diffusivity in our axisymmetric simulations, we developed a non-ideal version of our MHD code by adding diffusivity terms to our MHD equations. The diffusivity coefficient is proportional to the \citet{Shakura1973} $\alpha-$parameter. We take $\alpha=\alpha_d$ inside the radius $r < 5$ where the inner disc
interacts with the magnetosphere, and $\alpha=0$ at larger radii so that the MRI-driven accretion proceeds normally in the rest of the disc (high diffusivity damps the MRI). We also introduce the threshold density $\rho_d$ below which the diffusivity is not applied. Therefore, we take into account only the higher-density matter of the inner disc and exclude diffusive penetration of the coronal matter. This restriction helps to imitate the Rayleigh-Taylor instability at the
disc-magnetosphere boundary which develops more easily at sufficiently high densities \citep[this is because low-density small-scale filaments coalesce forming larger and denser filaments which penetrate through the magnetosphere, e.g.][]{Kulkarni2008}

\begin{figure*}
\centering
\includegraphics[width=1.0\textwidth]{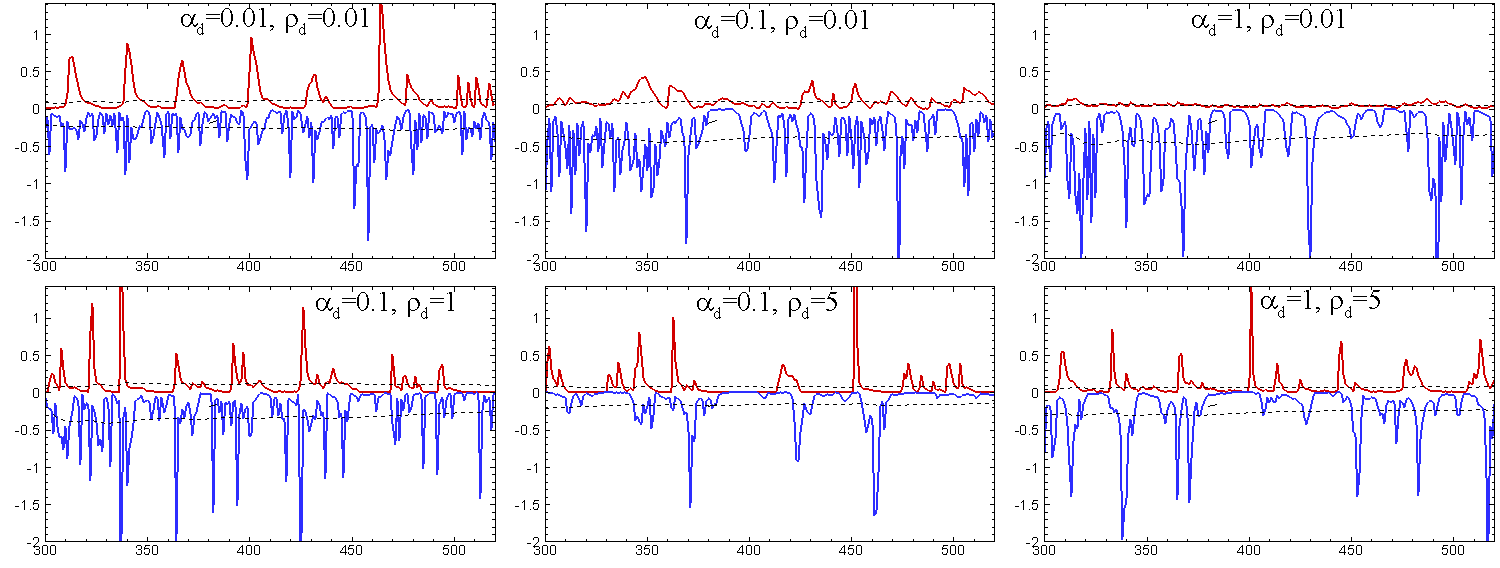}
\caption{Matter fluxes to the star (top curves) and to the wind (bottom curves),
calculated through the surface  $r=20$, $z=20$ and restricted by velocity $v>0.2$, for
different values of $\alpha_d$ and density thresholds $\rho_d$ (at which the
diffusivity is applied).} \label{fluxes-diff-6}
\end{figure*}

As a base, we take the case of the propelling star with corotation radius $r_{cor}=1.5$ and vary the diffusivity $\alpha_d$ as well as the matter density above which the diffusivity operates $\rho_d$. In the first series of runs, we choose a relatively small threshold density $\rho_d=0.01$ and perform simulations at three different diffusivities:  $\alpha_d=0.01, 0.1$ and $1$ (see set 1 in Table \ref{tab_diffusivity}). In another series of runs, we take large diffusivities: $\alpha_d=0.1$ or $1$ and high threshold densities: $\rho_d=1$ and  $\rho_d=5$ (set 2 in Table \ref{tab_diffusivity}). This second set of simulations is meant to study the Rayleigh-Taylor mechanism in the magnetosphere.

For each simulation, we calculate fluxes of matter onto the star $\mdot_*$ and into the wind $\mdot_{\rm wind}$. The fluxes into the wind are calculated through the surface $z=20, r=20$ for velocities $v_p>0.2$. The top panels of Fig. \ref{fig_diff4} show that at small threshold densities $\rho_d=0.01$, the flux onto the star systematically decreases with increasing $\alpha_d$ and becomes less spiky, whereas the matter flux to the wind becomes somewhat larger. This dependence of fluxes on diffusivity can be easily understood: at larger values of $\alpha_d$, matter diffuses more efficiently through the field lines and is ejected faster than it can accumulate at the disc-magnetosphere boundary. In cases of low diffusivity, the situation is opposite: the accreting matter does not efficiently diffuse through the field lines, instead accumulating at the disc-magnetosphere boundary where it eventually overflows onto the star. This episodic accumulation-overflow results in the spikes in accretion which are observed.

We calculate averaged fluxes onto the star and into the wind using Equation \ref{eqn_timeavg}. We expect that a higher diffusivities will result in a higher outflow efficiency and indeed, we find that the accretion rate onto the star $\langle \mdot_*\rangle$ systematically decreases with diffusivity in the first set of simulations. Matter flux to the wind $\langle\mdot_{\rm wind}\rangle$ increases for $\alpha_d=0.1$ then decreases again when the diffusivity becomes too high at $\alpha_d=1$. The angular momentum flux from the star, $\langle \dot L_\star\rangle$ (which determines the spin-down of the star), correspondingly increases, then decreases. It has a maximum value of $2.7$, which is only $1.7$ times larger than that in case of $\alpha_d=0$. Therefore, if additional diffusivity were to be included in our base models, the spin-down rate of the star may increase, but only by a factor of $\sim$2. As expected, the efficiency of the propeller ${\langle{\dot M}_{\rm wind}\rangle}/({\langle{\dot M}_{\rm wind}\rangle}+\langle{\dot M}_*\rangle)$ systematically increases with $\alpha_d$.


In our second set of simulations, we only enable the diffusivity if the local density is larger
than a density threshold value, $\rho_d$. In these cases, matter accumulates at the disc-magnetosphere boundary until it exceeds the threshold density of $\rho_d$; then it efficiently diffuses into the magnetosphere, with some fraction accreting onto the star and the remainder ejected into the wind. The bottom panels of Fig. \ref{fluxes-diff-6} show that the accretion and a significant number of the ejections proceed in spikes, reflecting the long episodes of matter accumulation at the disc-magnetosphere boundary. The timescale between major spikes is determined by the timescale of the matter accumulation---typically on the order of tens of dynamical timescales. This timescale depends on both of the parameters, $\alpha_d$ and $\rho_d$.

Analysis of this section shows that the larger diffusivities expected in more realistic three-dimensional simulations may lead to somewhat stronger outflows and shorter spin-down timescales of the propelling stars. These larger diffusivities model the mixing instabilities at the disc-magnetosphere boundary which we do not see in our 2.5D simulations. Our investigation shows that the difference between the cases with very low diffusivity and high diffusivity is not substantial, differing only by a factor of 2. The reason is that in both cases (for the same accretion rate in the disc) roughly the same amount of matter is ejected to winds due to propeller mechanism. In the less diffusive runs, the ejection occurs episodically in bursts whereas in the more diffusive runs the process is smoother. On average, however, the matter fluxes are comparable.

\section{Stresses in the MRI disc} \label{appen_stresses}
The matter accretion onto a star in the propeller regime is determined by the
turbulent stress in the disc. The integrated matter (subscript ``$m$'')
and magnetic field (subscript ``$f$'') stresses in the co-moving frame
are given by the vertically averaged angular momentum fluxes (Equation \ref{eqn_angmom}):
\begin{equation}
\langle{T_m}\rangle = \frac{1}{2H} \int dz \rho v_r v_\phi -
\langle{\rho v_r}\rangle \langle{v_\phi}\rangle
\end{equation}
and
\begin{equation}
\langle{T_f}\rangle = - \frac{1}{2H} \int dz \frac{B_r B_\phi}{4\pi}
\end{equation}
where
\begin{equation}
\langle{v_\phi}\rangle = \frac{1}{\Sigma} \int dz\rho v_\phi,
~~\langle{\rho v_r}\rangle = \frac{1}{2H} \int dz \rho v_r,
\end{equation}
$\Sigma = \int dz \rho$ is the disc surface density and $2H$ is the total
 thickness of the disc which we set using a density threshold. The matter and
 magnetic pressures are given by
\begin{equation}
\langle{P_m}\rangle = \frac{1}{2H} \int dz P,
~~~\langle{P_f}\rangle = \frac{1}{2H} \int dz \frac{\mathbf B_{tot}^2}{8\pi}.
\end{equation}
The standard $\alpha-$parameters associated with matter and magnetic
stresses are defined as the height-averaged stress divided by the height-averaged pressure
$$
\alpha_m = \frac{\langle{T_m}\rangle}{\langle{P_m}\rangle} , ~~~
\alpha_f = \frac{\langle{T_f}\rangle}{\langle{P_m}\rangle}.
$$
\begin{figure}
\centering
\includegraphics[width=.45\textwidth]{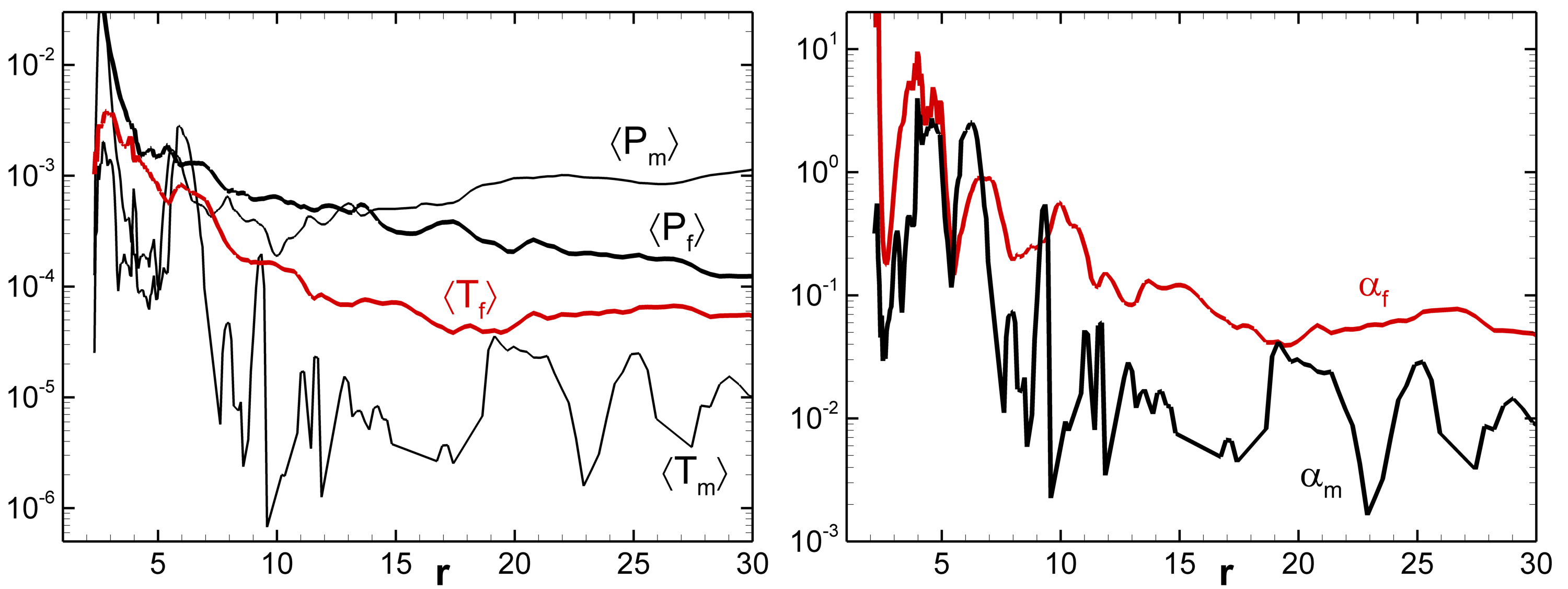}
\caption{{\bf Radial distribution of stresses in the disc.} The left panel
shows the $z$-averaged stresses and pressures inside the disc. Outside the
magnetospheric region, the magnetic stress dominates over the matter stress.
The right panel shows the computed $\alpha$-parameters.} \label{fig_stress}
\end{figure}
Fig. \ref{fig_stress} shows the radial distribution of stresses, pressure
and $\alpha-$parameters for the simulation. The magnetic stress $T_f$ roughly
an order of magnitude larger than the matter stress throughout the disc. Near
the magnetosphere, the magnetic pressure dominates over the matter pressure,
but is smaller in the external parts of the disc. Correspondingly, the magnetic
$\alpha$-parameter $\alpha_f$ is larger than the standard $\alpha$-parameter
determined by Reynolds stress, $\alpha_m$. In the magnetospheric region
($r\apprle 8$), both $\alpha$-parameters are very large because both
stresses are large. Further away from the star, we find $\alpha_f \approx 0.04-0.1$ and $\alpha_m \approx 0.001-0.08$ indicating that the radial angular momentum transport and inward accretion is determined by the magnetic stress.

\label{lastpage}

\end{document}